\documentclass{article}
\usepackage{amsmath,amssymb,amsfonts,hyperref}
\usepackage{graphicx,verbatim}
\usepackage{dsfont}

\usepackage{latexsym, amsmath, amsfonts, amssymb, amsthm, graphicx, epsfig, breqn, array, longtable, calc, multirow, hhline} 
\usepackage[lined,boxed,linesnumbered]{algorithm2e}

\begin{document}

\title{Optimal Design in Hierarchical Models with application in Multi-center Trials}
\author{Maryna Prus, Norbert Benda, Rainer Schwabe}
% Reihenfolge in Abhängigkeit von den Gepflogenheiten des Journals: In einem medizinisch-angewandten Journal wäre es korrekterweise  % Maryna (hauptverantwortlich), Norbert, Rainer (senior)
\maketitle

\begin{abstract}

Hierarchical random effect models are used for different purposes in clinical research and other areas. In general, the main focus is on population parameters related to the expected treatment effects or group differences among all units of an upper level (e.g. subjects in many settings). Optimal design for estimation of population parameters are well established for many models.  
However, optimal designs for the prediction for the individual units may be different. Several settings are identified in which individual prediction may be of interest. In this paper we determine optimal designs for the individual predictions, e.g. in multi-center trials, and compare them to a conventional balanced design with respect to treatment allocation.
Our investigations show, that balanced designs are far from optimal if the treatment effects vary strongly as compared to the residual error and more subjects should be recruited to the active (new) treatment in multi-center trials. 
Nevertheless, efficiency loss may be limited resulting in a moderate sample size increase when individual predictions are foreseen with a balanced allocation.

\end{abstract}

\section{Introduction}
Hierarchical random effect models are used for different purposes. Common applications, e.g. in clinical research, are random effect meta analyses of several clinical trials using individual patient data (IPD), multi-center trials assuming random center effects or mixed or random effect models for repeated measurements (MMRM) for longitudinal data within subjects. Clinical trials or meta-analyses of clinical trials may involve several populations or subgroups of patients (e.g. defined by genetic biomarkers), where in some settings between-population variability may be modeled using a hierarchical approach. Apart from clinical trials or observational studies data on quality parameters of drugs arise from hierarchical settings with multiple layers, i.e. with multiple sources of variability according to the underlying manufacturing process, e.g. given by manufacturing sites, production batches and samples within batches. In general, the main focus is on population parameters related to the expected treatment effects or group differences \emph{among} all units of an upper level (e.g. trials in IPD meta-analyses, centers in multi-center trials, patients in longitudinal trials, batches in quality control, etc.). Several authors considered optimal design for estimation of population parameters (expected values of random effects) in similar models, see, e.g. \cite{fedo}, \cite{fed2}, \cite{schm} and \cite{lem}.
% Schwabe, R. and Schmelter, T. (2008). On optimal designs in random intercept models. \textit{Tatra Mountains Mathematical Publications}. \textbf{39}, 145-153.
However, prediction of the outcome in the \emph{individual} units may also be of interest in several settings, as for treatment effects in single centers to assess qualification of individual clinics, manufacturing sites in manufacturing control, treatment effects in different subpopulations of patients. In these cases, the question arises, whether optimal designs for populations parameters can also be used for individual predictions of random effects, whether optimal designs for individual predictions differ from those for populations parameters and to which extent and which efficiency loss (or sample size increase) can be anticipated if another conventional design is chosen.

Therefore, we investigated optimal designs for random effects to be applied, e.g. in multi-center trials and compared them to a conventional balanced design with respect to treatment allocation.

The structure of the paper is as follows: 
In the second section a model of a multi-center trial is specified and the best linear unbiased predictions of the individual center parameters (intercepts and treatment effects) are derived. 
In Section~3 analytical results are presented to characterize optimal designs for prediction in multi-center trials. 
The results are illustrated by some numerical examples. 
The paper is concluded by a short discussion.

\section{Model Specification}\label{s2}
We consider a multicenter trial with $K$ different centers. 
In all of these centers individuals are allocated to two treatment groups. 
In the first group (denoted by $x=1$) the individuals receive an active treatment while in the second group (denoted by $x=0$) a placebo or a control treatment is applied. 

Denote by $\mu_i$ and $\alpha_i$ the intercept (mean response at placebo or control) and the effect of the active treatment (compared to placebo or control), respectively, in center $i=1,...,K$, which both may vary across the centers.
The response $Y_{ij}$ of an individual $j=1,...,N_i$ in center $i$ can be described as
\begin{equation}\label{mc}
{Y}_{ij}=\mu_{i}+\alpha_{i}x_{ij} + \varepsilon_{ij}\,,
\end{equation}
where $x_{ij}$ is equal to $1$, if the individual belongs to the treatment group, and $x_{ij}$ is equal to $0$ for the control (or placebo) group and $\varepsilon_{ij}$ denotes the random variation in the response of the individuals.
The individual variations $\varepsilon_{ij}$ are assumed ot have zero mean and to be homoscedastic with common variance $\sigma^2$.

The centers are assumed to be similar and, hence, representatives of a larger entity of centers. 
Thus the center specific intercepts $\mu_{i}$ and treatment effects $\alpha_{i}$ can be assumed as random with (unknown) expected values $\mathrm{E}(\mu_{i})=\mu$ and $\mathrm{E}(\alpha_{i})=\alpha$ characterizing the mean intercept and mean treatment effect across the centers and covariance structure $\mathrm{Cov}((\mu_i, \alpha_i)^\top)=\sigma^2\mathbf{D}$ for some $2\times 2$ positive definite dispersion matrix $\mathbf{D}$.
All random effects and all individual variations are assumed to be uncorrelated.

For the sake of simplicity we further assume that the total number $N$ of individuals is the same for all centers ($N_i=N$) and that the allocation rate is constant across the centers, i.\,e.\  the number $n$ of individuals in the treatment group is the same for all centers. 
The design problem can then be formulated in terms of finding the optimal allocation rate $w=n/N$ for the treatment group.
 
Because of exchangeability of the individuals within each center we may sort them in the analysis regardless of randomization in such a way that the first $n$ individuals $j=1,...,n$ to be analyzed receive the active treatment and the remaining $N-n$ individuals $j=n+1,...,N$ are in the control group.
Then the experimental settings $x_{ij}$ in \eqref{mc} can be specified by
\begin{equation*}
x_{ij}=x_j=\left\{\begin{array}{ll} 1, & j=1,..., n \\ 0, & j=n+1,..., N\end{array}\right.
\end{equation*}
and are independent of the center $i$.

Hence, the multi-center model \eqref{mc} can be identified as a particular case of the random coefficient regression model 
\begin{equation}\label{rcr}
Y_{ij}=\mathbf{f}(x_j)^\top\mbox{\boldmath{$\beta$}}_i+\varepsilon_{ij},\quad i=1, \dots, K, \quad j=1, \dots, N
\end{equation}
investigated by \cite{pru1}
when the regression functions and the center parameters are specified by $\mathbf{f}(x)=(1, x)^\top$ and
$\mbox{\boldmath{$\beta$}}_i=(\mu_i,\alpha_i)^\top$, respectively.
In general this model can be written in vector notation as
\begin{equation}\label{vec}
\mathbf{Y}_{i}=\mathbf{F}\mbox{\boldmath{$\beta$}}_i+\varepsilon_{i},
\end{equation}
where $\mathbf{Y}_{i}=(Y_{i1},...,Y_{iN})^\top$ and  $\varepsilon_{i}=(\varepsilon_{i1},...,\varepsilon_{iN})^\top$ are the $N$-dimensional vectors of observations and individual variations at center $i$, respectively, and $\mathbf{F}=(\mathbf{f}(x_1), ..., \mathbf{f}(x_N))^\top$ is the within center design matrix which is equal across all centers. 

For the present multicenter model the design matrix $\mathbf{F}$ simplifies to
\begin{equation*}
\mathbf{F}=\left(\begin{array}{cc}\mathds{1}_n & \mathds{1}_n\\ \mathds{1}_{N-n} & \mathbf{0}_{\,N-n}\end{array}\right),
\end{equation*}
where $\mathds{1}_\ell$ and $\mathbf{0}_{\,\ell}$ denote the $\ell$-dimensional vectors with all entries equal to $1$ and $0$, respectively. 

According to \cite{pru1} (see also \cite{fedo}) in model \eqref{rcr} the best linear unbiased predictors (BLUP) 
\begin{equation}\label{pred}
\hat{\mbox{\boldmath{$\beta $}}}_i=(\mathbf{F}^\top \mathbf{F}+\mathbf{D}^{-1})^{-1}(\mathbf{F}^\top \mathbf{F}\,\hat{\mbox{\boldmath{$\beta $}}}_{i;{\rm ind}}+\mathbf{D}^{-1}\hat{\mbox{\boldmath{$\beta $}}}_0)
\end{equation}
of the random parameters $\mbox{\boldmath{$\beta$}}_i$ are weighted combinations of the individual estimates  $\hat{\mbox{\boldmath{$\beta $}}}_{i;{\mathrm{ind}}}=(\mathbf{F}^\top \mathbf{F})^{-1}\mathbf{F}^\top \mathbf{Y}_i$ based only on the observations in  center $i$ and the the best linear unbiased estimator (BLUE)  $\hat{\mbox{\boldmath{$\beta $}}}_0=(\mathbf{F}^\top \mathbf{F})^{-1}\mathbf{F}^\top \bar{\mathbf{Y}}$ of the population parameter $\mbox{\boldmath{$\beta $}}_0=E(\mbox{\boldmath{$\beta $}}_i)$, where $\bar{\mathbf{Y}}=\frac{1}{K}\sum_{i=1}^{K}\mathbf{Y}_i$ is the mean observational vector averaged across the centers..

We additionally assume that the center intercepts $\mu_i$ and the center treatment effects $\alpha_i$ are uncorrelated for all centers, i.\,e.\ $\mathbf{D}=\mathrm{diag}(u,v)$, where $u=\sigma_{\mu}^2/\sigma^2>0$ and $v=\sigma_{\alpha}^2/\sigma^2>0$ are the variance ratios of the intercepts and the treatment effects in relation to the observational variance of the individuals. 

With the standard notations $Y_{i\,\cdot}^{(T)}=\frac{1}{n}\sum_{j=1}^nY_{ij}$ and $Y_{i\,\cdot}^{(C)}=\frac{1}{N-n}\sum_{j=n+1}^NY_{ij}$ for the mean response in the treatment (``$T$'') and the control (``$C$'') groups in center $i$, $Y_{\cdot\,\cdot}^{(T)}=\frac{1}{K}\sum_{i=1}^KY_{i\,\cdot}^{(T)}$ and $Y_{\cdot\,\cdot}^{(C)}=\frac{1}{K}\sum_{i=1}^KY_{i\,\cdot}^{(C)}$ for the overall mean of the treatment and the control groups, respectively, the BLUPs for the center parameters $\hat{\mbox{\boldmath{$\beta $}}}_i=(\hat{\mu}_i,\hat{\alpha}_i)^\top$ of the random intercepts and the random treatment effects ${\mbox{\boldmath{$\beta $}}}_i=({\mu}_i,{\alpha}_i)^\top$ in model \eqref{mc} can be written as weighted averages
\begin{equation}
\hat{\mu}_i=c_0(Y_{i\,\cdot}^{(T)}-Y_{\cdot\,\cdot}^{(T)})+cY_{i\,\cdot}^{(C)}+(1-c)Y_{\cdot\,\cdot}^{(C)},
\end{equation}
where $c_0=\frac{nu}{(Nu+1)(nv+1)-n^2uv}$ and $c=\frac{u(N-n)(nv+1)}{(Nu+1)(nv+1)-n^2uv}$,
and
\begin{equation}
\hat{\alpha}_i=c_1Y_{i\,\cdot}^{(T)}+(1-c_1)Y_{\cdot\,\cdot}^{(T)}-c_2Y_{i\,\cdot}^{(C)}-(1-c_2)Y_{\cdot\,\cdot}^{(C)}
\end{equation}
with weights $c_1=\frac{(Nu+1)nv-n^2uv}{(Nu+1)(nv+1)-n^2uv}$ and $c_2=\frac{Nu+nv+1}{(Nu+1)(nv+1)-n^2uv}$.

To measure the quality of a design we will use the mean squared error (MSE) matrix of the BLUP $\hat{\mbox{\boldmath{$\beta$}}}=\left(\hat{\mbox{\boldmath{$\beta$}}}_1^\top,..., \hat{\mbox{\boldmath{$\beta$}}}_K^\top\right)^\top$ of the complete vector $\mbox{\boldmath{$\beta$}}=(\mbox{\boldmath{$\beta$}}_1^\top,..., \mbox{\boldmath{$\beta$}}_K^\top)^\top$ of all random parameters. 
The mean squared error (MSE) matrix for the BLUP $\hat{\mbox{\boldmath{$\beta$}}}$ can be computed by means of the following formula :
\begin{eqnarray}
\mathrm{Cov}\left(\hat{\mbox{\boldmath{$\beta$}}}-\mbox{\boldmath{$\beta$}}\right)
&=& \sigma^2\left({\textstyle{\frac{1}{K}}}(\mathds{1}_K\mathds{1}_K^\top )\otimes  (\mathbf{F}^\top \mathbf{F})^{-1}\right.\nonumber \\
 && + \left.(\mathds{I}_K-{\textstyle{\frac{1}{K}}}\mathds{1}_{K}\mathds{1}_{K}^\top )\otimes (\mathbf{F}^\top \mathbf{F}+\mathbf{D}^{-1})^{-1} \right)
 \label{mse}
\end{eqnarray}
 (see \cite{pru1}), where $\mathds{I}_\ell$ denotes the $\ell\times \ell$ identity matrix and $\otimes$ is the symbol for the Kronecker product of matrices or vectors.

Further denote by $\mbox{\boldmath{$\Psi$}}_\alpha=\left(\mbox{\boldmath{$\alpha$}}_1,..., \mbox{\boldmath{$\alpha$}}_K\right)^\top$ the vector of treatment effects for all centers. 
Then $\mbox{\boldmath{$\Psi$}}_\alpha= \left(\mathbb{I}_K\otimes(0,\, 1) \right)\mbox{\boldmath{$\beta$}}$ and, hence, 
\begin{equation}
\mathrm{Cov}\left(\hat{\mbox{\boldmath{$\Psi$}}}_\alpha-\mbox{\boldmath{$\Psi$}}_\alpha\right)=\left(\mathbb{I}_K\otimes(0,\, 1) \right)\mathrm{Cov}\left(\hat{\mbox{\boldmath{$\beta$}}}-\mbox{\boldmath{$\beta$}}\right)\left(\mathbb{I}_K\otimes(0,\, 1) \right)^\top
\end{equation} 
for the MSE matrix of the BLUP $\hat{\mbox{\boldmath{$\Psi$}}}_\alpha=(\hat{\alpha}_1,..., \hat{\alpha}_K)^\top$ of $\mbox{\boldmath{$\Psi$}}_\alpha$.
Using this and formula \eqref{mse} we obtain the MSE matrix 
\begin{eqnarray}
\mathrm{Cov}\left(\hat{\mbox{\boldmath{$\Psi$}}}_\alpha-\mbox{\boldmath{$\Psi$}}_\alpha\right)&=&\sigma^2\left(\frac{N}{n(N-n)}\frac{1}{K}\mathds{1}_{K}\mathds{1}_{K}^\top\right.\nonumber\\
&&\mbox{}+\left.\frac{v(Nu+1)}{(Nu+1)(nv+1)-n^2uv}(\mathbb{I}_K-\frac{1}{K}\mathds{1}_{K}\mathds{1}_{K}^\top)\right).\label{mse1}
\end{eqnarray}

\section{Optimal Design}

As individuals are interchangeable within treatment groups we may define an exact within center design
\begin{equation}
\left(\begin{array}{cc} T & C \\ n & N-n \end{array}\right)
\end{equation}
by the allocation numbers $n$ and $N-n$ to the treatment and control group $T$ and $C$, respectively. 

For analytical purposes, we generalize this to the definition of an approximate design:
\begin{equation}\label{adesign}
\xi = \left(\begin{array}{cc} T & C \\ w & 1-w \end{array}\right),
\end{equation}
where $w=\frac{n}{N}$ is the allocation rate to the treatment group and $1-w=\frac{N-n}{N}$ is the allocation rate to the control group. 
For finding an optimal design only the optimal allocation rate $w^*$ to the treatment group has to be determined.

For an approximate design the definition of the MSE matrix \eqref{mse1} of the BLUP $\hat{\mbox{\boldmath{$\Psi$}}}_\alpha$ is extended in a straightforward manner an can be rewritten (neglecting $\sigma^2$) as :
\begin{eqnarray}
\mathrm{MSE}(w)&=&\frac{1}{Nw(1-w)}\frac{1}{K}\mathds{1}_{K}\mathds{1}_{K}^\top\nonumber \\
&&+\frac{v(Nu+1)}{(Nu+1)(Nwv+1)-N^2w^2uv}(\mathbb{I}_K-\frac{1}{K}\mathds{1}_{K}\mathds{1}_{K}^\top)\label{mse2}
\end{eqnarray}
 in terms of the allocation rate $w$.
 The approach of approximate designs seems in so far to be appropriate as the total number $N$ of individuals in each center should be sufficiently large.
 Otherwise optimal exact designs have to be obtained by adequate rounding of $w^*$ to a multiple of $1/N$ (see below).

For the assessment of the MSE matrix we focus on the $A$-optimality criterion which averages the mean squared errors of the center treatment effects.
More specifically, the $A$-criterion $\Phi_\alpha$ is the trace of the MSE matrix of the prediction $\hat{\mbox{\boldmath{$\Psi$}}}_\alpha$ of the center treatment effects. 
For an approximate design we get 
\begin{eqnarray}\label{a}
\Phi_\alpha(w)=\frac{1}{Nw(1-w)}+(K-1)\frac{v(Nu+1)}{(Nu+1)(Nwv+1)-N^2w^2uv}
\end{eqnarray}
for the criterion function $\Phi_\alpha$ in terms of the allocation rate $w$.
Because, in general, there is no explicit solution for the optimal allocation rate, which minimizes \eqref{a}, we will give an insight in the qualitative behavior by some numerical example below.

It is worth-while mentioning that the criterion \eqref{a} is convex and, therefore, an optimal exact design may be obtained by choosing the best of two exact designs adjacent to an optimal approximate design.

%Was ich nicht vestehe ist:
%
%Bayesian A (oder D)-Kriterium:
%
%\begin{eqnarray}
%\Phi_{Bayes}(n)=\frac{v(Nu+1)}{(Nu+1)(nv+1)-n^2uv}
%\end{eqnarray}
%
%Dann OD ist
%
%\begin{eqnarray}
%n^*_{Bayes}=\frac{N}{2}+\frac{1}{2u}
%\end{eqnarray}
%
%unabhängig von $v$ (Dispersion von $\alpha_i$) und monoton fallend in $u$...

\subsection{Example.}

For illustrative purposes we consider a numerical example with $K=50$ centers and $N=10$ individuals in each center. 
Figure~1 exhibits the behavior of the optimal allocation rate $w^*$ to the treatment group in dependence of the variance ratio $v$ of the treatment effects for some fixed values $0.01$, $0.1$, $0.25$, $0.5$ and $1,5$ of the variance ratio $u$ of the intercept. 
For reasons of presentation we plot the optimal allocation rate $w^*$ against the rescaled variance ratio $r_v=v/(1+v)$ in the spirit of intra-class correlation in order to cover all possible values of the treatment effects variance by a finite interval ($(0,1)$). 
Each value of the variance ratio $u$ of the intercepts is represented by one solid line.
What can be seen from the picture is that for fixed values of $u$  the optimal allocation rate $w^*$ is equal to $0.5$ for $v\to 0$ and increases with increasing values of the variance ratio $v$ of the treatment effects. 
The different lines associated to the different values of $u$ appear in descending order which means that the optimal allocation rates decrease when the variance ratio of the intercepts gets larger. 

The next figure (Figure~2) shows the behavior of the optimal allocation rate in dependence of the variance ratio $u$ of the intercepts for fixed values $0.01$, $0.1$, $0.2$, $0.5$ and $2$ of the variance ratio $v$ of the treatment effects, where again the variance ratio is rescaled ($r_u=u/(u+1)$). 
Also here it can be seen that the optimal allocation rate decreases with increasing values of $u$ and increases with increasing values of $v$.
    \begin{figure}[ht]
    \begin{minipage}[]{6.6cm}
       \centering
       \includegraphics[width=6.6cm]{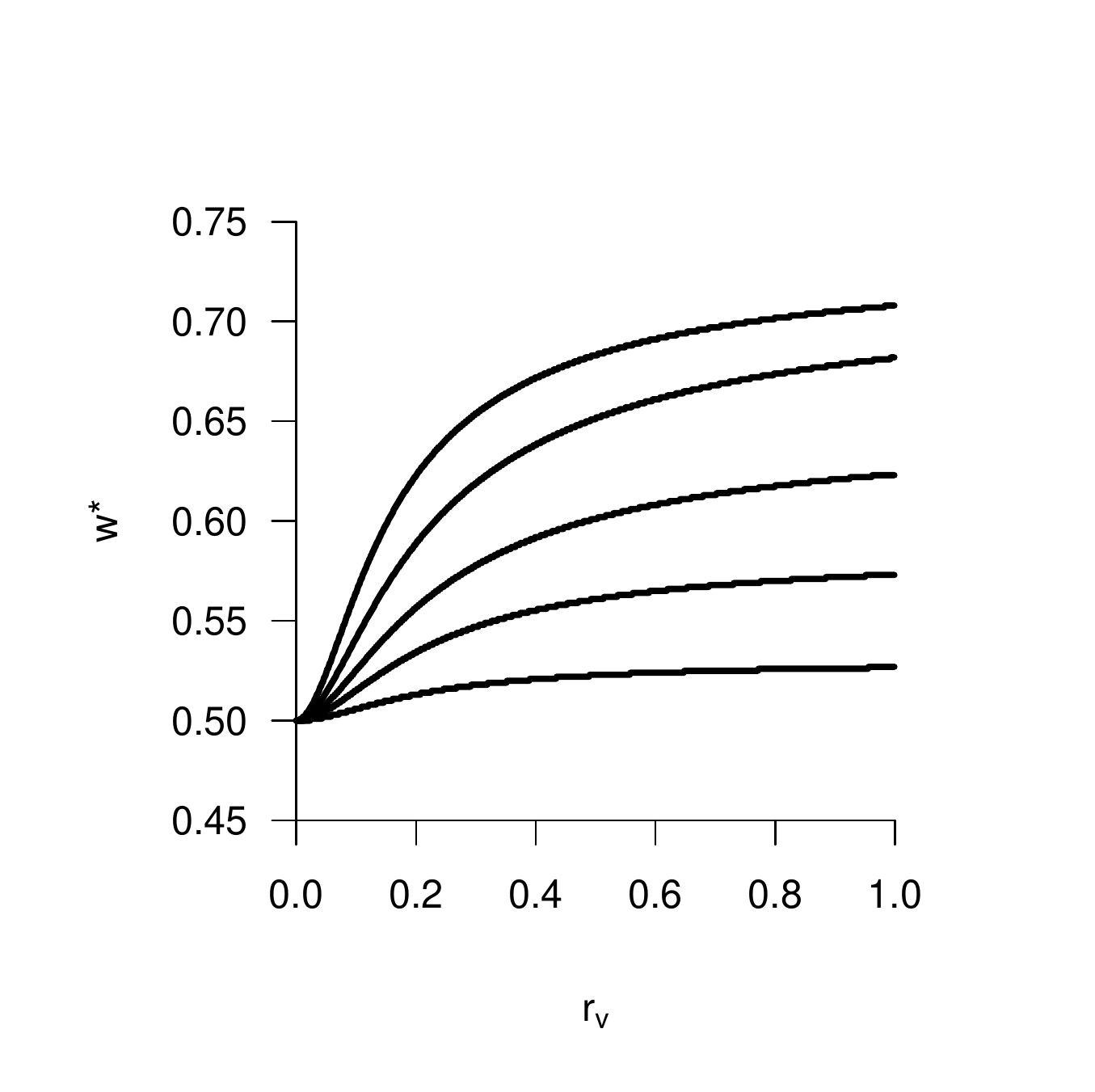}
       \end{minipage}
       \begin{minipage}[]{6.6cm}
       \centering
       \includegraphics[width=6.6cm]{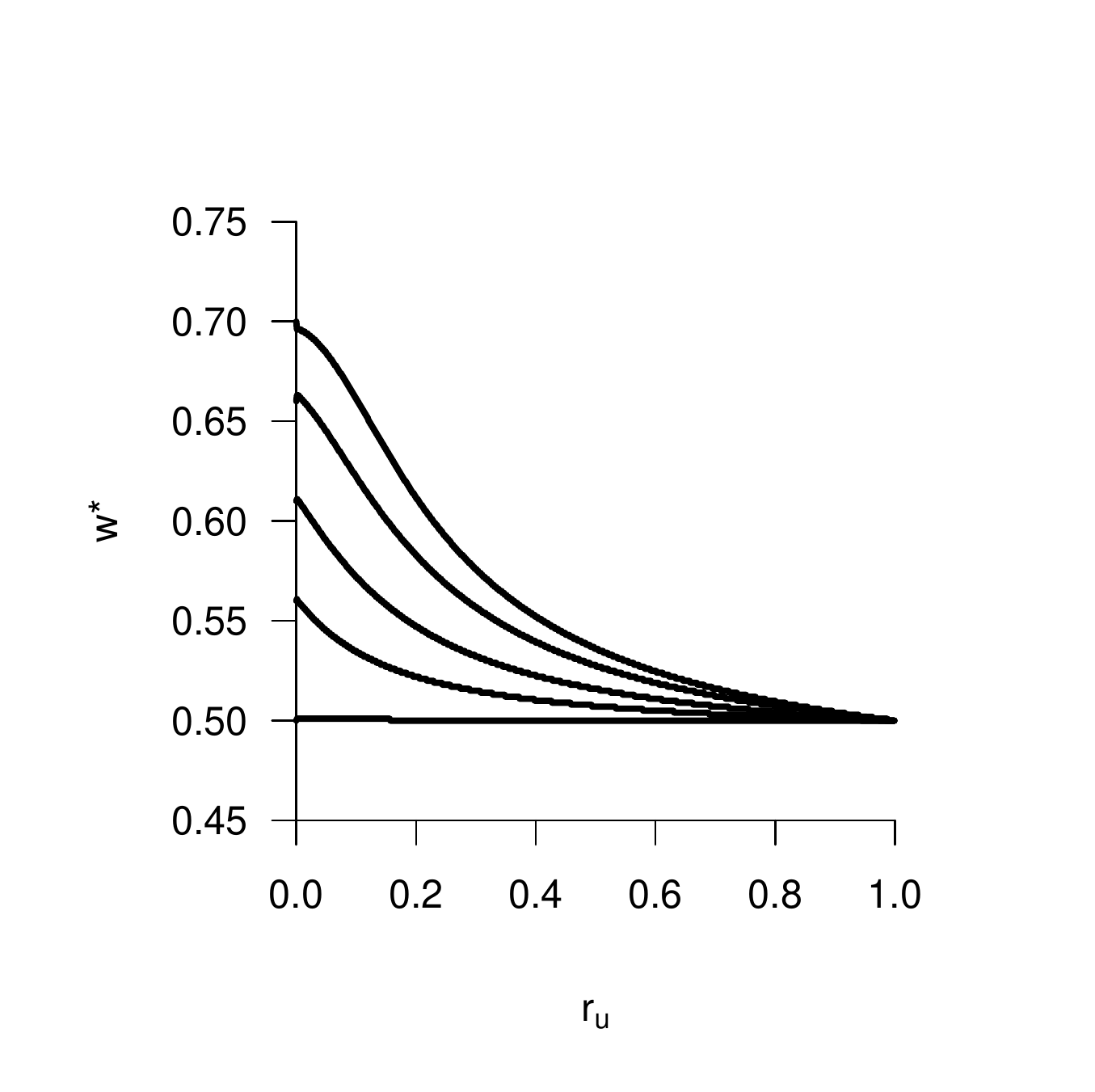}
       \end{minipage}
       \vspace{-1mm}
       \hspace*{5 mm}
       \begin{minipage}[]{6.3cm}
       Figure 1: \textit{A}-optimal allocation rate $w^*$ to the treatment group in dependence on the rescaled variance ratio $r_v=v/(1+v)$ of the treatment effects for $u=0.01$, $0.1$, $0.25$, $0.5$ and $1.5$
       \end{minipage}
       \hspace{6 mm}
       \begin{minipage}[]{6.3cm}
              Figure 2: \textit{A}-optimal allocation rate $w^*$ to the treatment group in dependence on the rescaled variance ratio $r_u=u/(1+u)$ of the intercepts for $v=0.01$, $0.1$, $0.2$, $0.5$ and $2$
       \end{minipage}
    \end{figure}   
		
Finally, Figures~3 and 4 present the efficiency of the equal allocation rate $w_0=0.5$ which is optimal in the fixed effects model ($u=v=0$). The efficiency for the $A$-criterion  ($A$-efficiency) has been computed using the standard formula
\begin{equation}
\mathrm{eff}(w_0)=\frac{\Phi_\alpha(w^*)}{\Phi_\alpha(w_0)}.
\end{equation}
		    \begin{figure}[ht]
    \begin{minipage}[]{6.6cm}
       \centering
       \includegraphics[width=6.6cm]{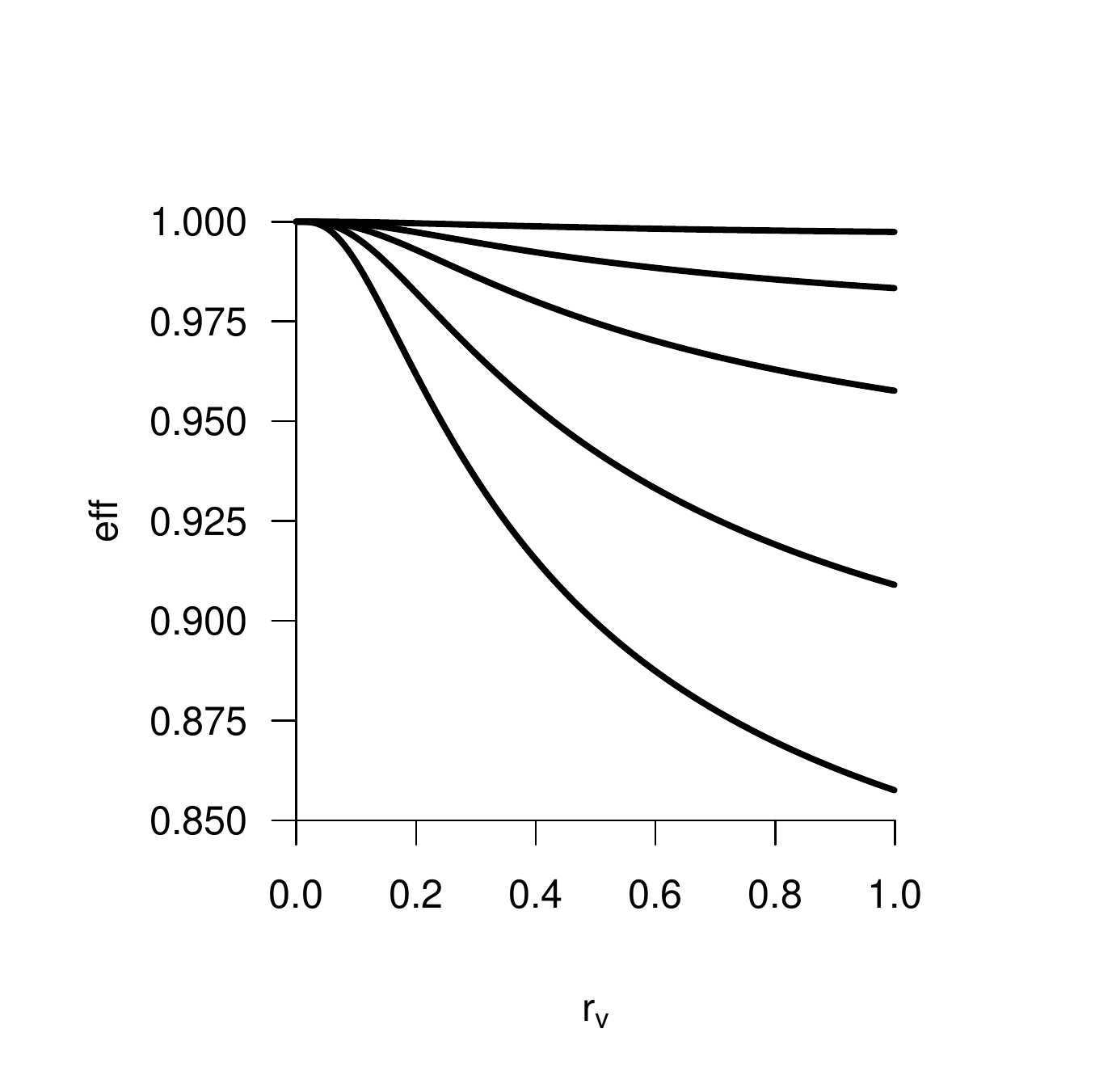}
       \end{minipage}
       \begin{minipage}[]{6.6cm}
       \centering
       \includegraphics[width=6.6cm]{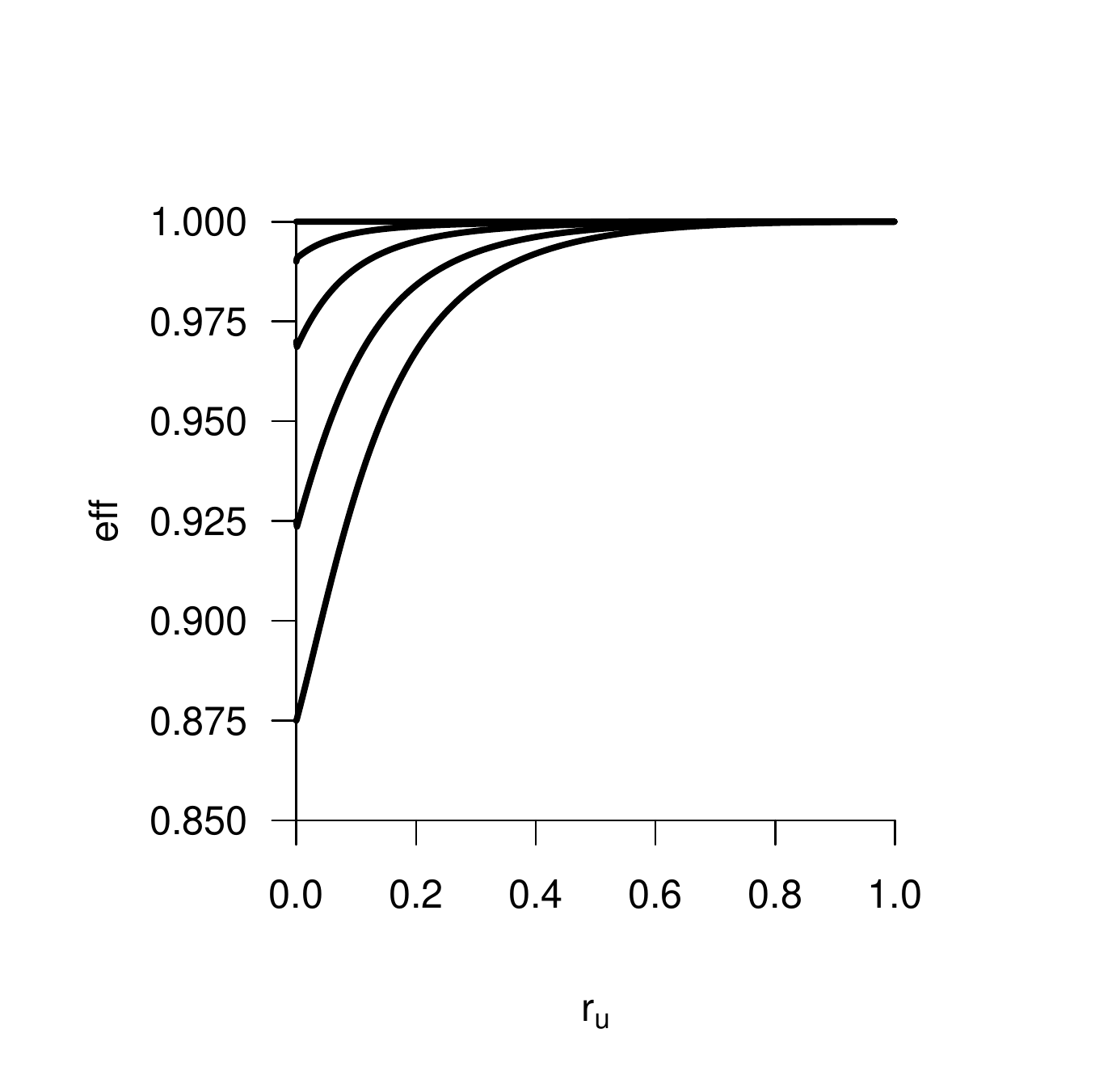}
       \end{minipage}
       \vspace{-1mm}
       \hspace*{5 mm}
       \begin{minipage}[]{6.3cm}
       Figure 3: $A$-efficiency of the equal allocation rate $w_0=0.5$ in dependence on the rescaled variance ratio  $r_v=v/(1+v)$ of the treatment effects for $u=0.01$, $0.1$, $0.25$, $0.5$ and $1.5$
       \end{minipage}
       \hspace{6 mm}
       \begin{minipage}[]{6.3cm}
       Figure 4: $A$-efficiency of the equal allocation rate $w_0=0.5$ in dependence on the rescaled variance ratio  $r_u=u/(1+u)$ of the intercepts for $v=0.01$, $0.1$, $0.2$, $0.5$ and $2$
       \end{minipage}
    \end{figure}
		The efficiencies decrease with increasing values of the variance of the treatment effects and increase with increasing values of the variance of the intercepts.
		
		\textbf{Example cont.}

For illustrative purposes we consider numerical examples with $K=50$ or $K=100$ centers and $N=10$ or $N=5$ individuals in each center. 
Figures~5 and 6 exhibit the behavior of the optimal allocation rate $w^*$ to the treatment group in dependence of the variance ratio $q=v/u$ for some fixed values $0.01$ (solid line), $0.1$ (dashed line), $0.25$ (dotted line), $0.5$ (dashed-dotted line) and $1,5$ (dashed line, long dashes) of the variance ratio $u$ of the intercept. 
For reasons of presentation we plot the optimal allocation rate $w^*$ against the rescaled variance ratio $r_q=q/(1+q)$ in the spirit of intraclass correlation in order to cover all possible values of variance ratio by a finite interval ($(0,1)$). As we can observe on the graphics, the dependence on the variance ratio is more essential for $K=100$ and $N=5$ than for $K=50$ and $N=10$.
    \begin{figure}[ht]
    \begin{minipage}[]{6.6cm}
       \centering
       \includegraphics[width=6.6cm]{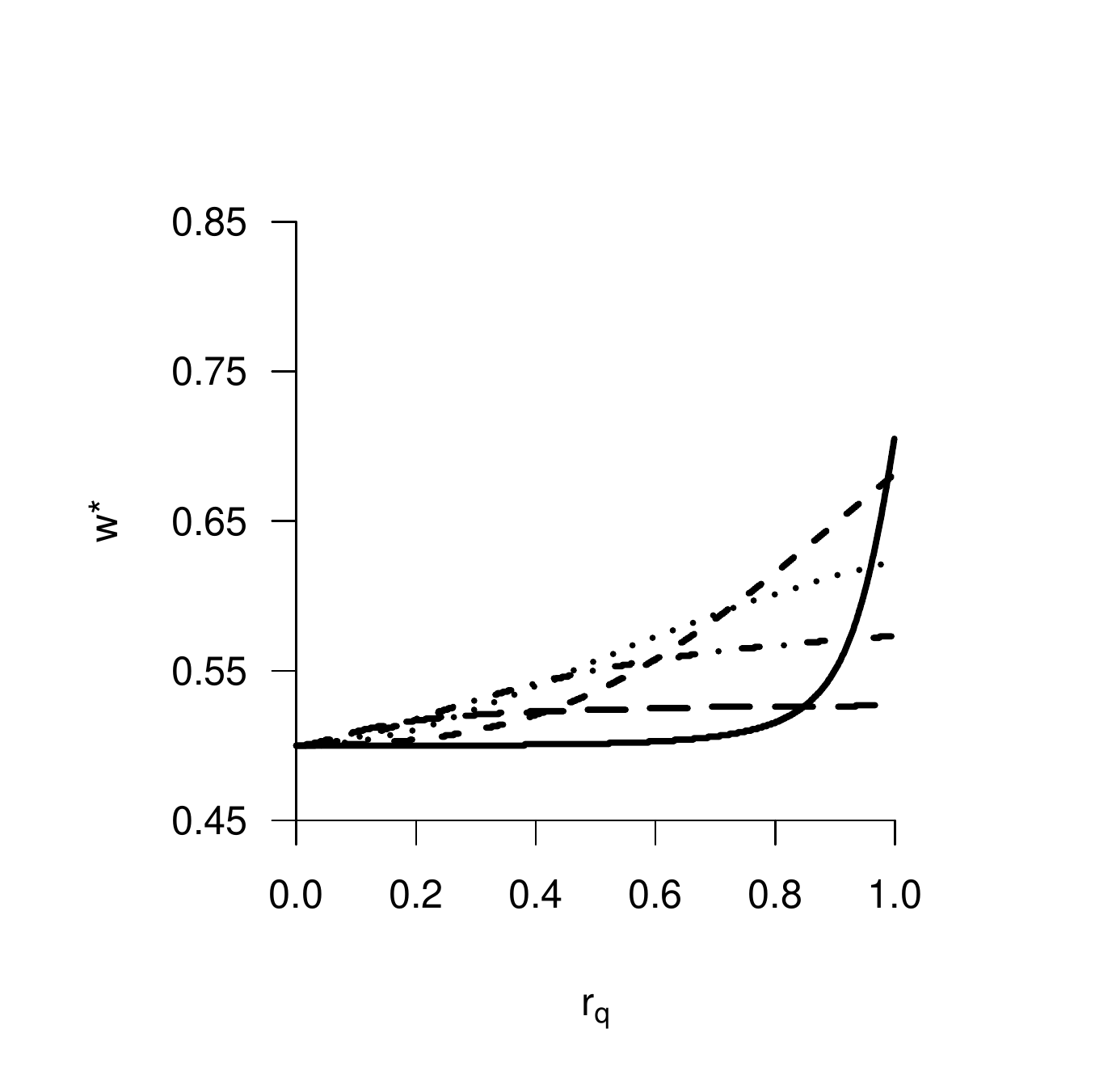}
       \end{minipage}
       \begin{minipage}[]{6.6cm}
       \centering
       \includegraphics[width=6.6cm]{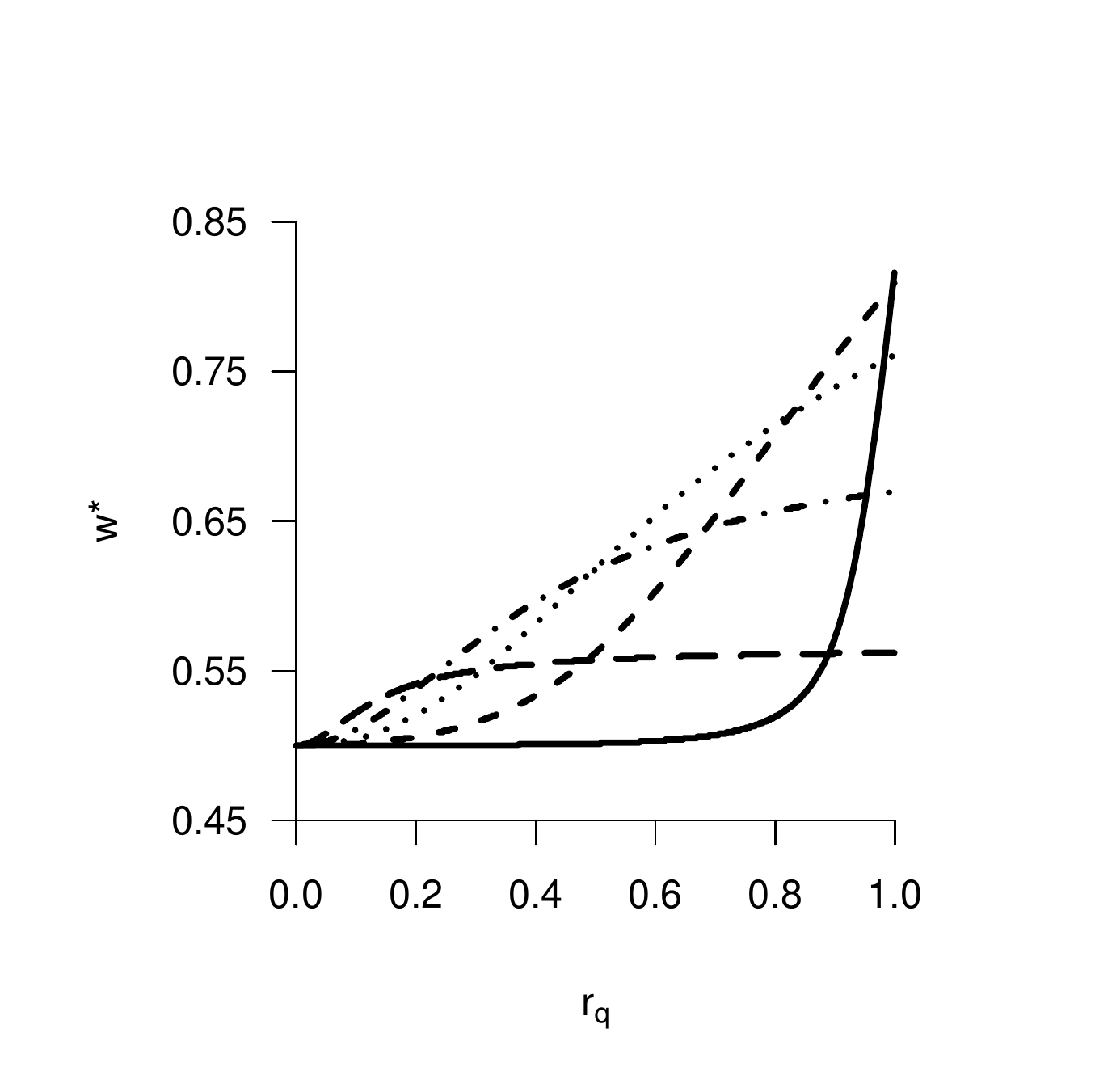}
       \end{minipage}
       \vspace{-1mm}
       \hspace*{5 mm}
       \begin{minipage}[]{6.3cm}
       Figure 5: \textit{A}-optimal allocation rate $w^*$ to the treatment group in dependence on the rescaled variance ratio $r_q=q/(1+q)$ of the treatment effects for $u=0.01$, $0.1$, $0.25$, $0.5$, $1.5$, $K=50$ and $N=10$
       \end{minipage}
       \hspace{6 mm}
       \begin{minipage}[]{6.3cm}
              Figure 6: \textit{A}-optimal allocation rate $w^*$ to the treatment group in dependence on the rescaled variance ratio $r_q=q/(1+q)$ of the treatment effects for $u=0.01$, $0.1$, $0.25$, $0.5$, $1.5$, $K=100$ and $N=5$
       \end{minipage}
    \end{figure}   
		
Figures~7 and 8 presents the efficiency of the equal allocation rate $w_0=0.5$ which is optimal in the fixed effects model. 
The efficiency is more sensible with respect to the variance ratio in the case $K=100$ and $N=5$ than for $K=50$ and $N=10$.
		    \begin{figure}[ht]
    \begin{minipage}[]{6.6cm}
       \centering
       \includegraphics[width=6.6cm]{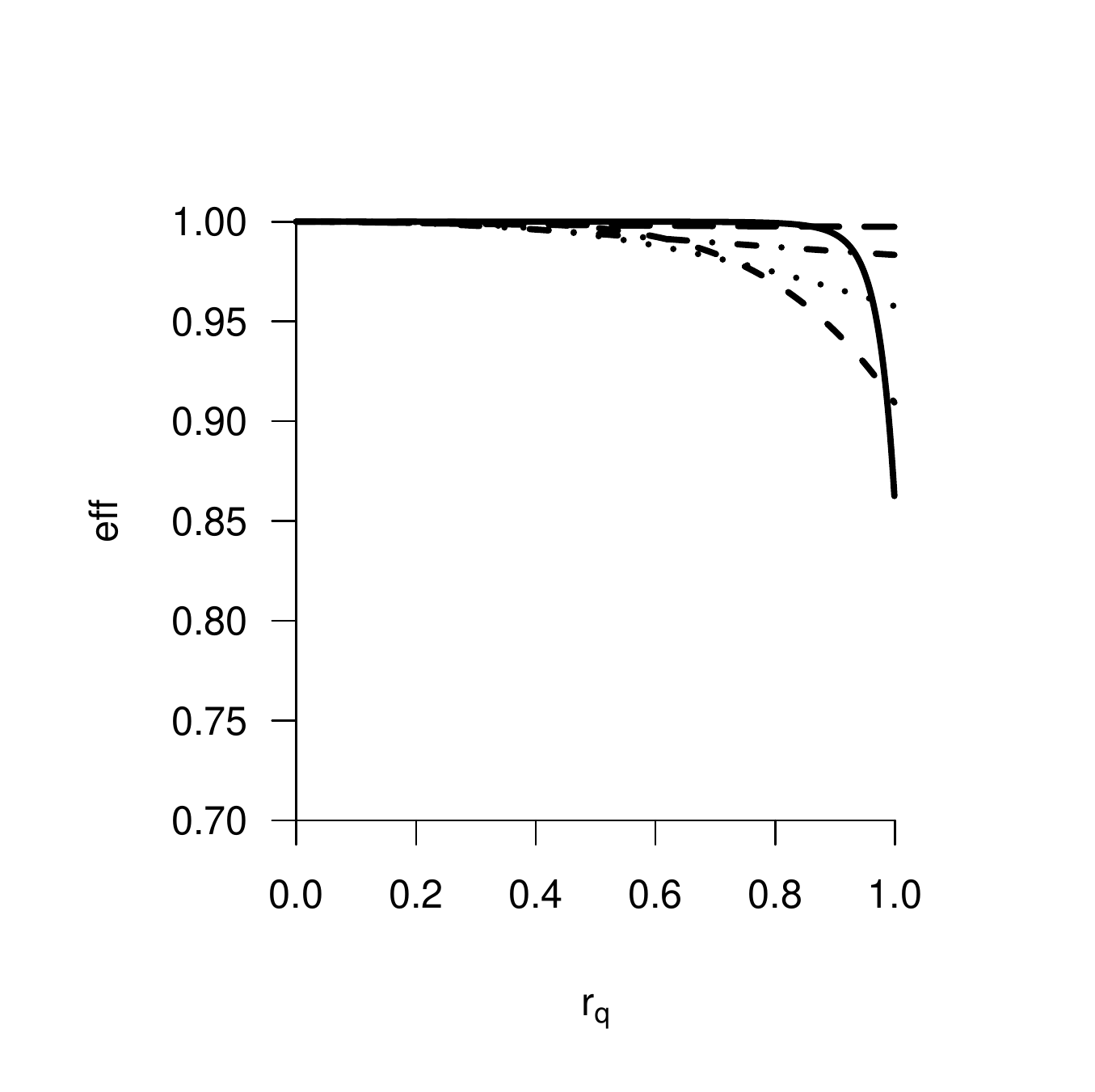}
       \end{minipage}
       \begin{minipage}[]{6.6cm}
       \centering
       \includegraphics[width=6.6cm]{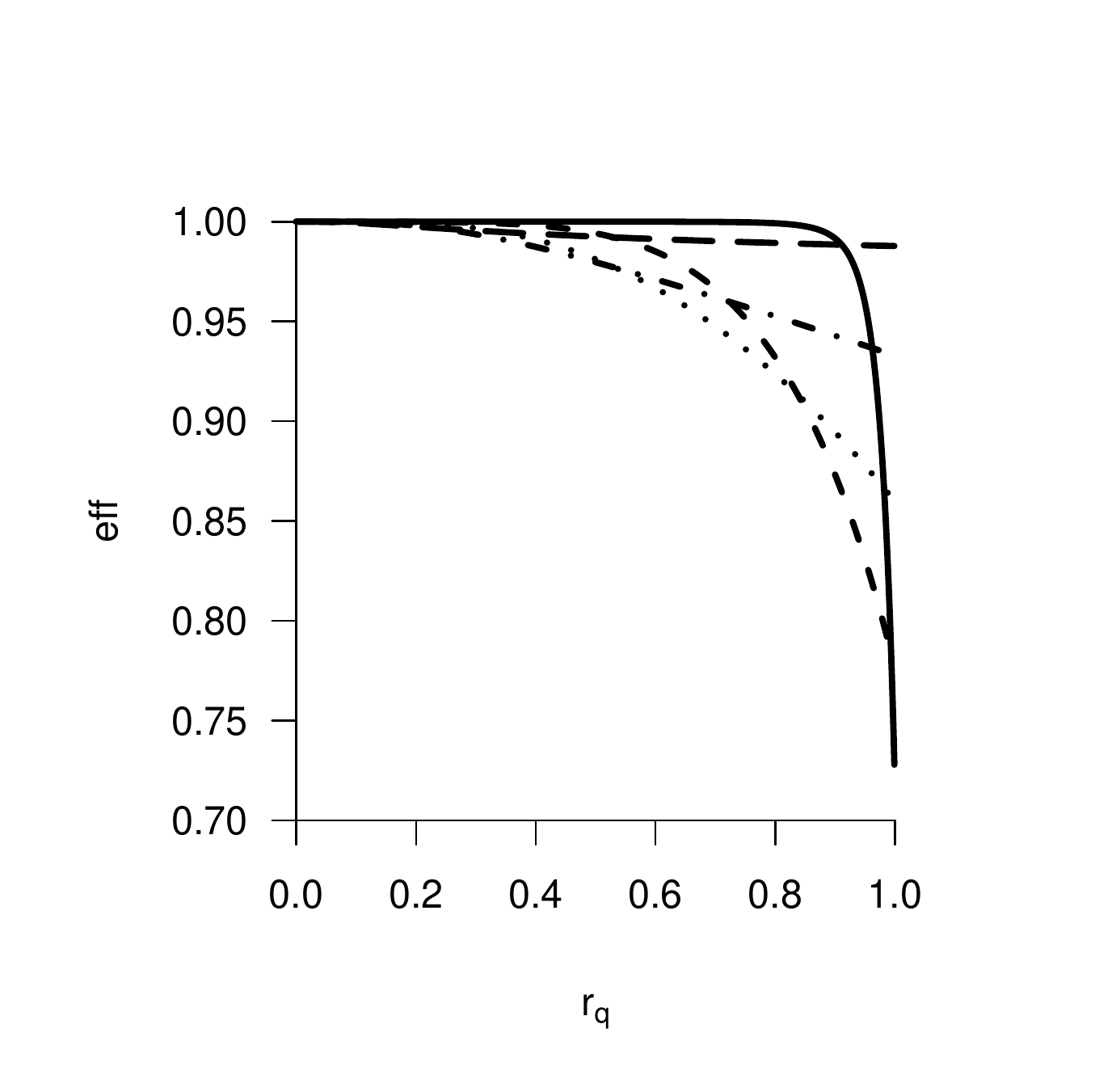}
       \end{minipage}
       \vspace{-1mm}
       \hspace*{5 mm}
       \begin{minipage}[]{6.3cm}
       Figure 7: $A$-efficiency of the equal allocation rate $w_0=0.5$ in dependence on the rescaled variance ratio  $r_q=q/(1+q)$ of the intercepts for $v=0.01$, $0.1$, $0.25$, $0.5$, $1.5$, $K=50$ and $N=10$
       \end{minipage}
       \hspace{6 mm}
       \begin{minipage}[]{6.3cm}
       Figure 8: $A$-efficiency of the equal allocation rate $w_0=0.5$ in dependence on the rescaled variance ratio  $r_q=q/(1+q)$ of the intercepts for $v=0.01$, $0.1$, $0.25$, $0.5$, $1.5$, $K=100$ and $N=5$
       \end{minipage}
    \end{figure}
		
\section{Discussion}
As illustrated in the example, the larger the between-unit (between-trial) variability of the treatment effects the more the optimal weight deviates from equal allocation, especially if the variance of the units' intercepts is small. An increasing heterogeneity in the treatment effect leads to a decreased precision of the design that is optimal for population parameters: A balanced design is far from optimal if the treatment effects vary strongly as compared to the residual error and more subjects should be recruited to the active (new) treatment in multi-center trials. Nevertheless, it appears reassuring to the clinical trial practitioner that the efficiency loss may be limited as in the example resulting in a total sample size increase of about 10 - 20 \% in the considered scenarios if individual predictions are foreseen with a balanced allocation. Usually, between-unit variability of treatment effects are considered to be rather small, indicating that equal allocation may suffice. However, using the results given in the paper, specific settings with different expectations can be assessed properly, in order to make optimally use of a limited number of patients or sample units to predict random effects of individual units.

\bibliographystyle{natbib}
\bibliography{prus}

\end{document}